\documentclass[11pt]{article}
\usepackage[english]{babel}

\usepackage{amssymb}
\usepackage{graphicx}
\usepackage[T1]{fontenc}
\begin{document}

\title{Matter power spectrum for the generalized Chaplygin gas model: The relativistic case}
\author{J. C. Fabris\footnote{e-mail: fabris@pq.cnpq.br}, H.E.S. Velten\footnote{velten@daad-alumni.de. Present address:
Fakult\"at f\"ur Physik, Universit\"at Bielefeld, Postfach 100131, 33501 Bielefeld, Germany.} and W. Zimdahl\footnote{winfried.zimdahl@pq.cnpq.br}\\
Departamento de F\'{\i}sica, Universidade Federal do Esp\'{\i}rito Santo, \\
CEP 29060-900 Vit\'{o}ria, Esp\'{\i}rito Santo, Brasil
}
\date{}
\maketitle

\begin{abstract}
The generalized Chaplygin gas (GCG) model is the prototype of a unified model of dark energy (DE) and dark
matter (DM). It is characterized by equation-of-state (EoS) parameters $A$ and $\alpha$. We use a statistical analysis of the 2dFGRS data to constrain these parameters.
In particular, we find that very small (close to zero) and very large values ($\alpha\gg 1$) of the equation-of-state parameter $\alpha$ are preferred.
To test the validity of this type of unification of the dark sector we admit the existence of a separate DM component in addition to the Chaplygin gas and calculate the probability distribution for the fractional contributions of both components to the total energy density. This analysis favors a model for which the Universe is nearly entirely made up of the separate DM component with an almost negligible Chaplygin gas part. This confirms the results of a previous Newtonian analysis.
\end{abstract}

\vspace{0.5cm} \leftline{PACS: 98.80.-k, 04.62.+v}

\section{Introduction}

The Chaplygin gas model represents the best known proposal of a unification of DM and DE into a single fluid \cite{moschella}. In its generalized form \cite{berto}, the Chaplygin
gas is a fluid with an equation of state $p_{c} = - A/\rho_{c}^{\alpha}$, where $p_{c}$ is the pressure and $\rho_{c}$ is the energy density. Solving the
conservation equation
$\dot\rho_{c} + 3 \dot{a}/a \left(\rho_{c} + p_{c}\right) = 0$,
where $a$ is the scale factor of the spatially flat Robertson-Walker metric, the result for the energy density is
\begin{equation}
\rho_{c} = \left[A + B a^{-3(1 + \alpha}\right]^{1/\left(1+\alpha\right)}\ .
\label{rho}
\end{equation}
The present value of the scale factor is normalized to $a_{0} =1$. Combined with Friedmann's equation and assuming $\alpha > -1$, the solution (\ref{rho}) interpolates between a matter
phase for $a \ll 1$  and a de Sitter phase for $a \gg 1$.
In this sense, the GCG can play the roles
of dark matter in the past and of dark energy at present and future times, thus unifying the two components of the dark sector of the cosmic substratum into a single component. The original Chaplygin gas corresponds to $\alpha = 1$. It can be traced back to the Nambu-Goto action in the light cone parametrization \cite{jackiw}, that is, it can be seen as rooted in a string theoretical context. The generalization to $\alpha \neq 1$ is phenomenological \cite{berto}. However, the GCG  model has the advantage of providing an interpolation between the $\Lambda$CDM model, represented by $B=0$ and $\alpha = 0$, and other models, like, e.g., viscous models \cite{Szydlowski,BVM,avelino}. Similar approaches are the
so-called Cardassian models \cite{freese,feng}. Thus, the GCG model covers a large variety of dark energy models.
\par
A great deal of effort has been made in order to constrain the different parameters of the GCG model, mainly the parameter $\alpha$.
This includes data from supernova type Ia \cite{colistete}, the anisotropy of the cosmic microwave background \cite{amendola}, baryonic acoustic oscillations \cite{Wu}, the integrated Sachs-Wolfe effect \cite{tommaso} and the matter power spectrum (see \cite{neo} and references therein).
Constraints from combined data sources have been 
obtained in \cite{campo} and \cite{park}.

The supernova type Ia data were shown to be well described by the GCG gas, which seems to be favorable for the unified model. But the situation changes if the model is confronted with matter power spectrum data.
In this brief communication we generalize our previous Newtonian analysis \cite{neo} of the 2dFGRS program \cite{2d} to the fully relativistic case. We consider three configurations of a Chaplygin-gas cosmology.

\noindent (i) At first we assume the material content of the Universe to consist of a GCG, radiation and a pressureless baryon component with a fraction of $\Omega_{m0} = \Omega_{b0} = 0.043$, as suggested by the five-year WMAP data. In other words, we prescribe a unified model of the dark sector. The result is that, together with very small values of $\alpha$, for which the GCG behavior is close to that of a cosmological constant, also values $\alpha >2$ have a high probability. This confirms both an earlier qualitative but gauge invariant analysis of matter power spectrum data \cite{starobinsky} and our previous simplified Newtonian analysis \cite{neo}. Still more surprising is that very large values, i.e. values of several hundreds for $\alpha$, are preferred by the data.  This confirms an independent investigation of the integrated Sachs-Wolfe effect with preferred values up to $\alpha \approx 350$ \cite{piattella}.

\noindent (ii) At second we consider the original Chaplygin gas, i.e., the case $\alpha = 1$, but we leave the matter fraction free, thus admitting that the matter component is not just made up by the baryons. This is equivalent to allow for a separate DM component in addition to the GCG.
For this reason we write the fractional density parameter for the pressureless matter today as $\Omega_{m0}=0.043+\Omega_{dm0}$. This additional freedom is used to test the unified model of the dark sector. The unified model can be regarded as favored by the data if the probability distribution function (PDF) for the matter fraction is large around the value that characterizes the baryon fraction. If, on the other hand, the PDF is largest at a substantially higher value, the unified model has to be regarded as disfavored. Our fully relativistic study confirms the results of a previous Newtonian analysis according to which the matter fraction does not only not peak around the value for baryons, but it is maximal at values of $\Omega_{m0}$ close to $\Omega_{m0} = 1$. In other words, the data seem to prefer a model which is close to the Einstein-de Sitter universe. This contradicts the results from the SN Ia data according to which the unification scenario, i.e, a universe filled almost exclusively by the Chaplygin gas component, is favored \cite{colistete}.

\noindent (iii) At third we extend the analysis of (ii) to $\alpha\neq 1$. As far as the value of $\Omega_{m0}$ is concerned, we recover the result of (ii). The behavior of $\alpha$ is similar to that described under (i), although here it is less important since the GCG is subdominant.

Strictly speaking, models with $\alpha > 1$ seem to be un-physical since they correspond to a superluminal sound speed. However, with certain modification causality can be preserved \cite{starobinsky}.
To see the influence of a large $\alpha$ on the background dynamics of a GCG cosmology, we consider the EoS parameter
\begin{equation}
\frac{p_{c}}{\rho_{c}} = - \frac{\bar{A}}{\bar{A} + \left(1 - \bar{A}\right)a^{-3\left(1+\alpha\right)}} \qquad \mathrm{with} \qquad
\bar{A} = \frac{A}{\rho_{c,0}^{\alpha + 1}}\ .
\label{eos}
\end{equation}
The parameter $\alpha$ influences the value of $a$ at which the transition from decelerated to accelerated expansion occurs.
Assuming a unified model and denoting the transition value of the scale factor by $a_{q}$, one has
\begin{equation}
\frac{p_{c}}{\rho_{c}}|_{q} = - \frac{1}{3 }= - \frac{\bar{A}}{\bar{A} + \left(1 - \bar{A}\right)a_{q}^{-3\left(1+\alpha\right)}}\ .
\label{aqdef}
\end{equation}
Solving for $a_{q}$ yields
\begin{equation}
a_{q} = \left(\frac{1 - \bar{A}}{2\bar{A}}\right)^{\frac{1}{3(1+\alpha)}}\ .
\label{aqsol}
\end{equation}
Since $\frac{1 - \bar{A}}{2\bar{A}} < 1$ this means, that for growing values of $\alpha$ the transition
period $a_{q}$ approaches $a=1$, i.e., the present epoch. For very large $\alpha$, the matter period is longer and the transition to accelerated expansion occurs suddenly and more recently than in the $\Lambda$CDM model. This property has also been discussed in \cite{piattella}.

\section{Basic set of equations}

Our starting point are Einstein's equations coupled to a pressureless fluid, radiation and to the GCG. They read,
\begin{eqnarray}
R_{\mu\nu} &=& 8\pi G\biggr\{T^m_{\mu\nu} - \frac{1}{2}g_{\mu\nu}T^m\biggl\} + 8\pi G\biggr\{T^r_{\mu\nu} - \frac{1}{2}g_{\mu\nu}T^r\biggl\} + 8\pi G\biggr\{T^c_{\mu\nu} - \frac{1}{2}g_{\mu\nu}T^c\biggl\},\nonumber\\
{T_m^{\mu\nu}}_{;\mu} &=& 0 \quad , \quad {T_c^{\mu\nu}}_{;\mu} = 0 \quad , \quad {T_r^{\mu\nu}}_{;\mu} = 0 \nonumber
\end{eqnarray}
The superscripts (subscripts) $m$, $r$ and $c$ stand for "matter", "radiation" and "Chaplygin". We assume a perfect fluid structure for the cosmic medium as a whole and also for each of the components,
\begin{equation}
T^{\mu\nu} = \rho u^{\mu}u^{\nu} + p \left(g^{\mu\nu} - u^{\mu}u^{\nu}\right)\ ,
 \qquad T_{A}^{\mu\nu} = \rho_{A} u_A^{\mu} u^{\nu}_{A} + p_{A} \left(g^{\mu\nu} - u_{A}^{\mu}u_{A}^{\nu}\right) \
,\qquad A = m, c, r\ .
\label{T}
\end{equation}
Using now the flat Friedmann-Robertson-Walker metric
\begin{eqnarray}
ds^2 = dt^2 - a(t)^2[dx^2 + dy^2 + dz^2]\ ,\nonumber
\end{eqnarray}
and identifying all the background $4$-velocities,
Einstein's equations reduce to
\begin{eqnarray}
\biggr(\frac{\dot a}{a}\biggl)^2 &=& \frac{8\pi G}{3}\rho_m + \frac{8\pi G}{3}\rho_r + \frac{8\pi G}{3}\rho_c,\\
2\frac{\ddot a}{a} + \biggr(\frac{\dot a}{a}\biggl)^2 &=& - 8\pi Gp_c,\\
\dot\rho_m + 3\frac{\dot a}\rho_m = 0 \quad &\Rightarrow& \quad \rho_m = \rho_{m0}a^{-3},\\
\dot\rho_r + 4\frac{\dot a}\rho_r = 0 \quad &\Rightarrow& \quad \rho_r = \rho_{r0}a^{-4},\\
\dot\rho_c + 3\frac{\dot a}{a}(\rho_c + p_c) = 0 \quad (p_c = - A/\rho_c^\alpha) \quad &\Rightarrow& \quad \rho_c = \biggr\{A + \frac{B}{a^{3(1 + \alpha)}}\biggl\}^{1/(1 + \alpha)} .
\end{eqnarray}
\par
The perturbed equations in the synchronous gauge take the form,
\begin{eqnarray}
\frac{\ddot{h}}{2}+\frac{\dot{a}}{a}\dot{h}-4\pi G\left(\delta \rho+3\delta \,p\right) &=& 0\\
\dot{\delta \rho}+\frac{3\dot{a}}{a}\left(\delta\rho+\delta\,p\right)+\left(\rho+p\right)\left(\theta-\frac{\dot{h}}{2}\right) &=& 0,\\
\left(p+\rho\right)\dot{\theta}+\left[\left(\dot{\rho}+\dot{p}\right)+\frac{5\dot{a}}{a}\left(\rho+p\right)\right]\theta+\frac{\nabla^{2}\delta\,p}{a^{2}} &=& 0,
\end{eqnarray}
where $\rho$ and $p$ stand for the total matter and pressure, respectively, and $\theta = \delta u^{i}_{,i}$.
\par
In terms of the components, we end up with the following equations:
\begin{eqnarray}
\frac{\ddot{h}}{2}+\frac{\dot{a}}{a}\dot{h}-4\pi G\left[\delta \rho_{m}+\delta \rho_{c}+\delta \rho_{r}+3 (\delta \,p_{m}+\delta p_{c}+\delta \,p_{r})\right] &=& 0,\\
\dot{\delta \rho_{m}}+\frac{3\dot{a}}{a}\left(\delta\rho_{m}+\delta\,p_{m}\right)+\left(\rho_{m}+p_{m}\right)\left(\theta_{m}-\frac{\dot{h}}{2}\right) &=& 0,\\
\left(\rho_{m}+p_{m}\right)\dot{\theta_{m}}+\left[\left(\dot{\rho_{m}}+\dot{p_{m}}\right)+\frac{5\dot{a}}{a}\left(\rho_{m}+p_{m}\right)\right]\theta_{m}+\frac{\nabla^{2}\delta\,p_{m}}{a^{2}} &=& 0,\\
\dot{\delta \rho_{c}}+\frac{3\dot{a}}{a}\left(\delta\rho_{c}+\delta\,p_{c}\right)+\left(\rho_{c}+p_{c}\right)\left(\theta_{c}-\frac{\dot{h}}{2}\right) &=& 0,\\
\left(\rho_{c}+p_{c}\right)\dot{\theta_{c}}+\left[\left(\dot{\rho_{c}}+\dot{p_{c}}\right)+\frac{5\dot{a}}{a}\left(\rho_{c}+p_{c}\right)\right]\theta_{c}+\frac{\nabla^{2}\delta\,p_{c}}{a^{2}} &=& 0,\\
\dot{\delta \rho_{r}}+\frac{3\dot{a}}{a}\left(\delta\rho_{r}+\delta\,p_{r}\right)+\left(\rho_{r}+p_{r}\right)\left(\theta_{r}-\frac{\dot{h}}{2}\right) &=& 0,\\
\left(\rho_{r}+p_{r}\right)\dot{\theta_{r}}+\left[\left(\dot{\rho_{r}}+\dot{p_{r}}\right)+\frac{5\dot{a}}{a}\left(\rho_{r}+p_{r}\right)\right]\theta_{r}+\frac{\nabla^{2}\delta\,p_{r}}{a^{2}} &=& 0 \ ,
\end{eqnarray}
with $\theta_{m} = \delta u_{m,i}^{i}$, $\theta_{c} = \delta u_{c,i}^{i}$ and $\theta_{r} = \delta u_{r,i}^{i}$.
\par
With the definitions
\begin{eqnarray}
h(a) &=& \left(\bar{A}+\frac{1-\bar{A}}{a^{3\left(1+\alpha\right)}}\right)^{\frac{1}{1+\alpha}},\\
\Omega_{c}(a) &=& \Omega_{c0}h(a),\\
w(a)&=& -\frac{\bar{A}}{\left[h\left(a\right)\right]^{1+\alpha}},\\
v^{2}_{s}(a) &=& -\alpha w(a),\\
g(a)= \ddot a &=& -\frac{\Omega_{m0}}{2a^{2}}-\frac{\Omega_{c}\left(a\right)\left[1+3w\left(a\right)\right]}{2}-\frac{\Omega_{r0}}{a^{3}},\\
f(a)= \dot a^2 &=& -\frac{\Omega_{m0}}{a}-\Omega_{c}\left(a\right)+\frac{\Omega_{r0}}{a^{2}},
\end{eqnarray}
the set of first-order equations becomes
\begin{eqnarray}
\delta^{\prime\prime}+\left(\frac{g(a)}{f(a)}+\frac{2}{a}\right)\delta^{\prime}
-\frac{3\Omega_{m0}}{2a^{3}f(a)}\delta = \frac{3\Omega_{c}(a)}{2f(a)}\lambda\left[1+3v_{s}^{2}(a)\right]&+&\frac{3\Omega_{r0}}{a^{4}f(a)}\delta_{r}(a);
\\
\lambda^{\prime}+\frac{3}{a}\left[v_{s}\left(a\right)-w\left(a\right)\right]\lambda(a)+
\left(1+w(a)\right)\left[\frac{\theta_{c}(a)}{\sqrt{f(a)}}-\delta^{\prime}\right] &=& 0;\\
(1+w(a))\left[\theta_{c}^{\prime}+\frac{\left[2-3v_{s}^{2}(a)\right]}{a}\theta_{c}\right] &=&v^{2}_{s}(a)\left(\frac{k}{k_{0}}\right)^{2}\frac{\lambda}{\sqrt{f(a)}a^{2}};\\
\delta_{r}^{\prime}+\frac{4}{3}\left(\frac{\theta_{r}}{\sqrt{f}}-\delta^{\prime}\right) &=& 0;\\
\theta_{r}^{\prime}+\frac{\theta_r}{a} &=& \left(\frac{k}{k_{0}}\right)^{2}\frac{\delta_{r}}{4f(a)a^{2}}\ ,
\end{eqnarray}
where
\begin{equation}
\delta \equiv \frac{\delta\rho_{m}}{\rho_{m}}\ , \quad \lambda \equiv \frac{\delta\rho_{c}}{\rho_{c}}\ ,\quad\delta_{r} \equiv \frac{\delta\rho_{r}}{\rho_{r}}
\
\label{}
\end{equation}
and $k_{0}^{-1}=3000\,h\,Mpc$.

\section{Bayesian analysis}

\begin{center}
\begin{figure}[!t]
\begin{minipage}[t]{0.3\linewidth}
\includegraphics[width=\linewidth]{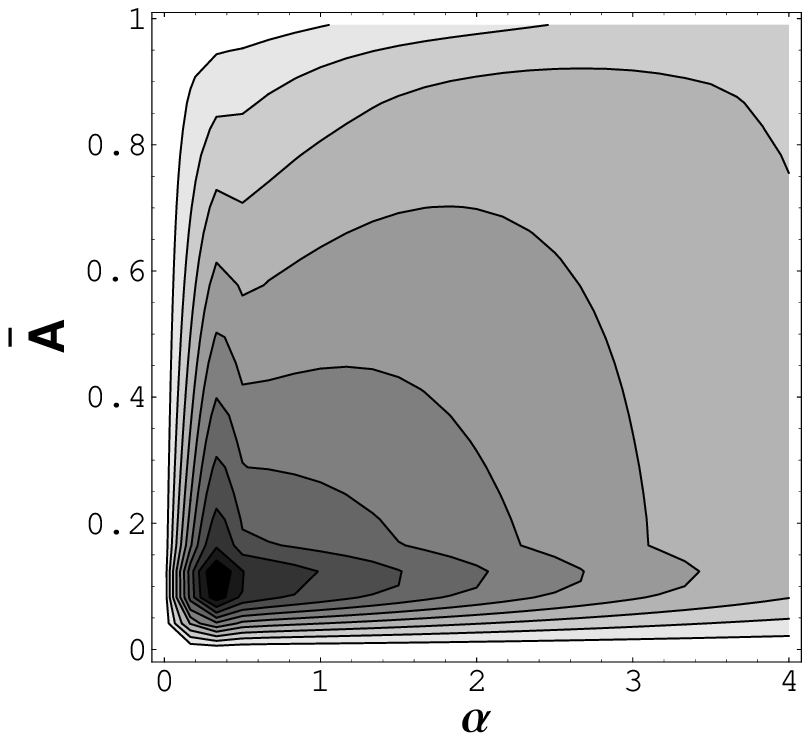}
\end{minipage} \hfill
\begin{minipage}[t]{0.3\linewidth}
\includegraphics[width=\linewidth]{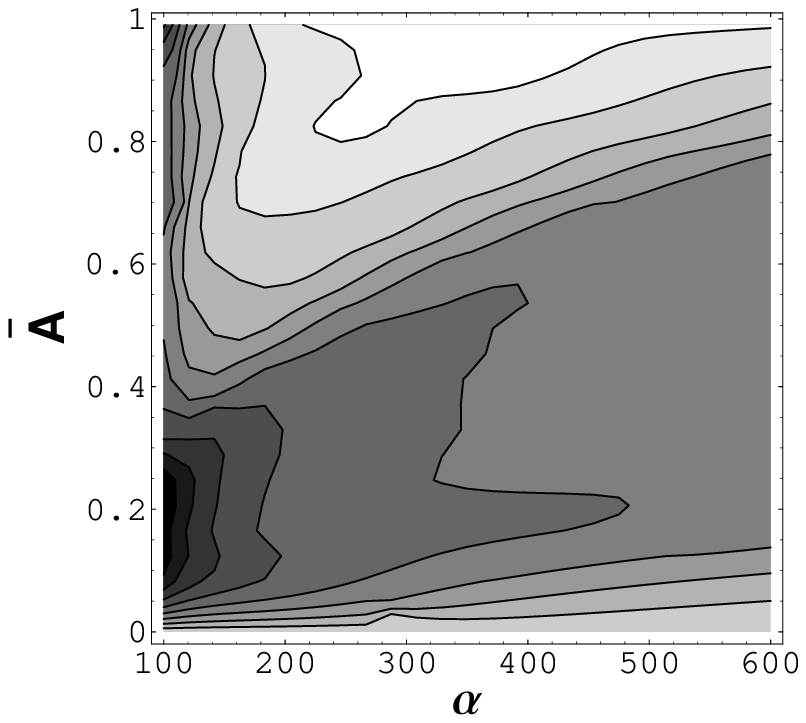}
\end{minipage} \hfill
\begin{minipage}[t]{0.3\linewidth}
\includegraphics[width=\linewidth]{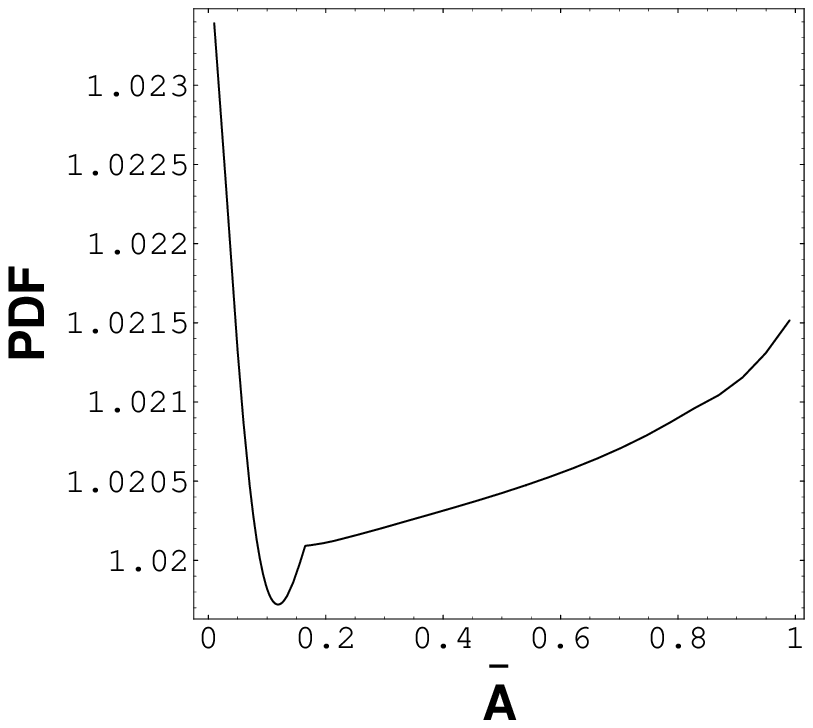}
\end{minipage} \hfill
\begin{minipage}[t]{0.3\linewidth}
\includegraphics[width=\linewidth]{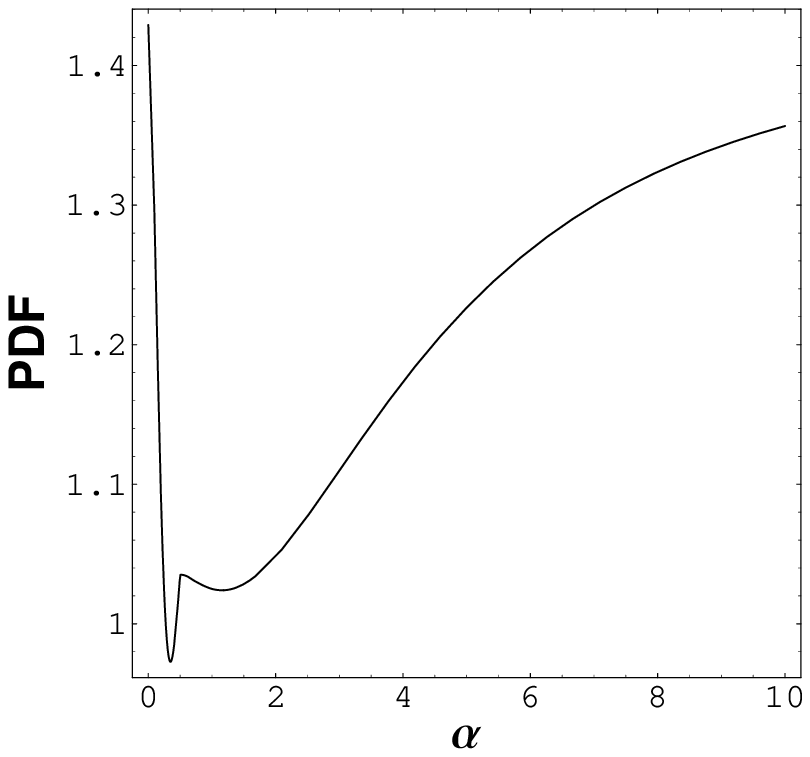}
\end{minipage} \hfill
\begin{minipage}[t]{0.3\linewidth}
\includegraphics[width=\linewidth]{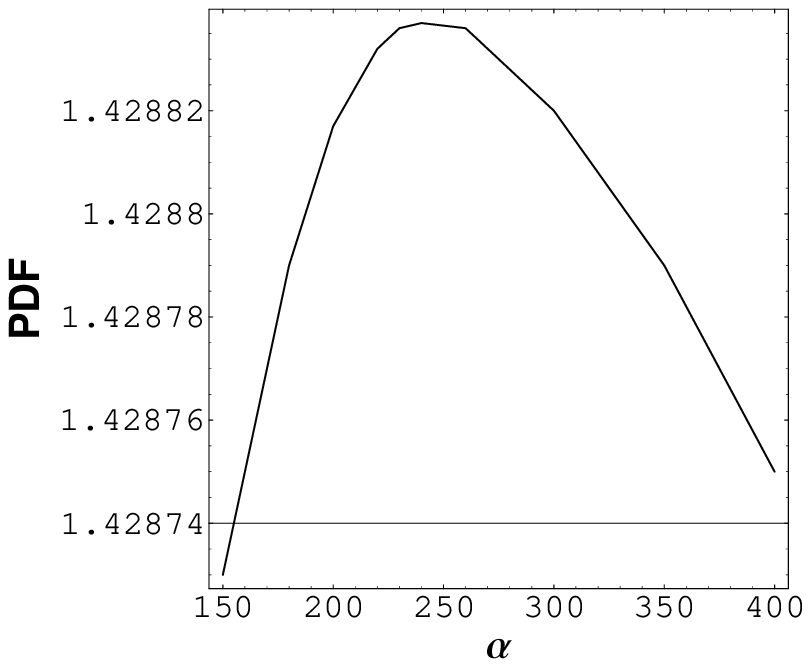}
\end{minipage} \hfill
\caption{{\protect\footnotesize PDFs for the unification scenario ($\Omega_{m0} = \Omega_{b0} =0.043$).
}}
\label{unific}
\end{figure}
\end{center}

The matter power spectrum is defined by
\begin{equation}
{\cal P} = \delta_k^2 \quad ,
\end{equation}
where $\delta_k$ is the Fourier transform of the dimensionless density
contrast $\delta$. We will constrain the free parameters using the quantity
\begin{equation}
\chi^2 = \sum_i\biggr(\frac{{\cal P}_i^o - {\cal P}_i^t}{\sigma_i}\biggl)^2
\quad ,
\label{chi}
\end{equation}
where ${\cal P}_i^o$ is the observational value for the power
spectrum, ${\cal P}_i^t$ is the corresponding theoretical result and
$\sigma_i$ denotes the error bar. The index $i$ refers to a measurement
corresponding to given wavenumber. The quantity (\ref{chi}) qualifies
the fitting of the observational data for a given theoretical
model with specific values of the free parameters.
Hence, $\chi^2$ is a function of the free parameters
of the model. The probability distribution function is then
defined as
\begin{equation}
F(x_n) = F_0\,e^{-\chi^2(x_n)/2} \quad ,
\end{equation}
where the $x_n$ denote the ensemble of free parameters and $F_0$
is a normalization constant.
In order to obtain an estimation for a given parameter one has to
integrate (marginalize) over all the other ones. For a more
detailed description of this statistical analysis see Ref.
\cite{colistete}. 
In the evaluation of the probability distribution, we will use the total $\chi^2$.
With the prior of a spatially flat universe, the analysis has to take into account three free parameters: $\alpha$, $\Omega_{m0}$ and $\bar A$
(or equivalently $\Omega_{dm0}$) and $\bar{A}$).
As already mentioned, we will consider three cases. For case (i), the unified model, the fraction $\Omega_{m0}$ is fixed to $\Omega_{m0} = \Omega_{b0} = 0.043$ and there are only two free parameters, $\alpha$ and $\bar A$. The PDFs are
visualized in figure \ref{unific}. The upper left and central panels show the two-dimensional PDF for $\bar A$ and $\alpha$ for different ranges of $\alpha$. The lighter the color the higher the probability. In the upper right panel the one-dimensional PDF for $\bar{A}$ is depicted. It has maxima close to zero and close to unity. Recall that $\bar A = 0$ means a pressureless medium while $\bar A = 1$ corresponds to a cosmological constant for any value of $\alpha$. Both lower panels show the one-dimensional PDFs for different ranges of $\alpha$. There is a maximum at $\alpha = 0$ which, in the background, corresponds to the $\Lambda$CDM model. Also values $\alpha > 2$ have a high probability. This feature is both in agreement with a previous gauge-invariant perturbation analysis \cite{starobinsky} and with our simplified Newtonian model \cite{neo}.
The more surprising result, however, is the existence of a maximum at very large values of $\alpha$, as can be seen in the lower right panel of figure \ref{unific}. This backs up an independent study of the integrated Sachs-Wolfe effect \cite{piattella} with values of $\alpha$ up to $350$. Here, we obtain a maximum for $\alpha = 240$.

\begin{center}
\begin{figure}[!t]
\begin{minipage}[t]{0.3\linewidth}
\includegraphics[width=\linewidth]{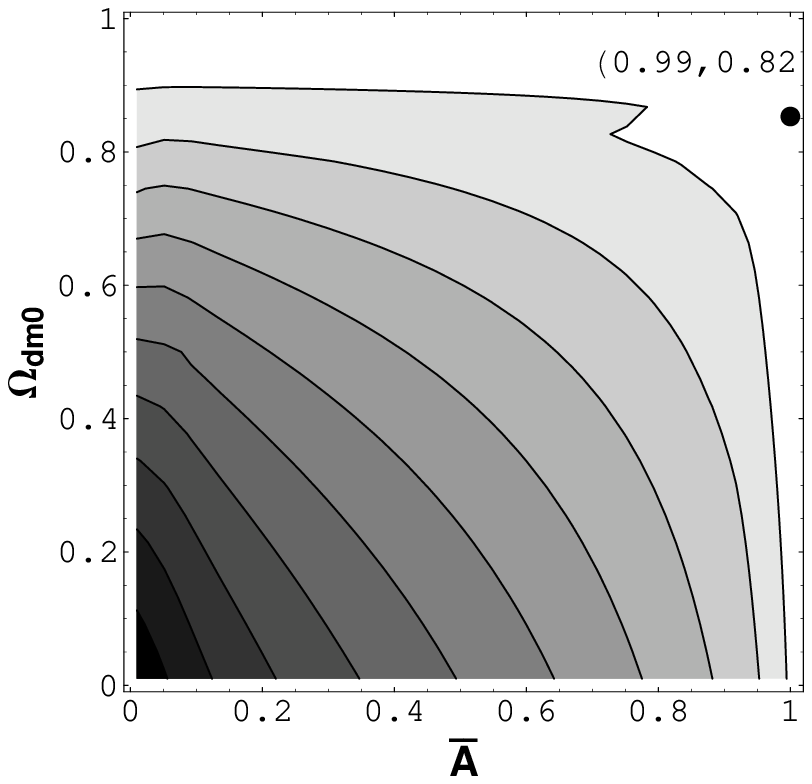}
\end{minipage} \hfill
\begin{minipage}[t]{0.3\linewidth}
\includegraphics[width=\linewidth]{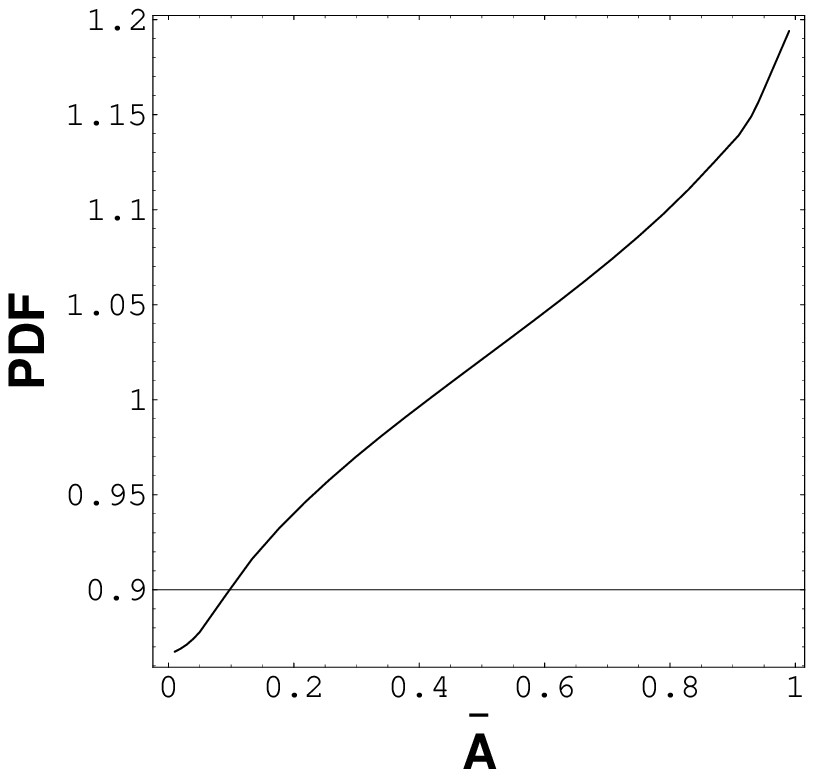}
\end{minipage} \hfill
\begin{minipage}[t]{0.3\linewidth}
\includegraphics[width=\linewidth]{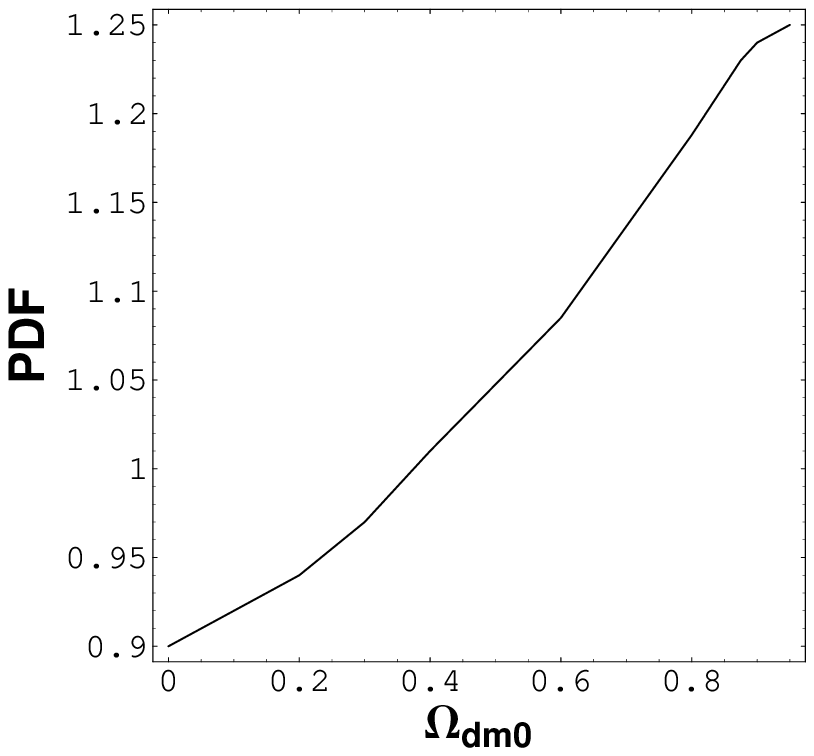}
\end{minipage} \hfill
\caption{{\protect\footnotesize PDFs for the original Chaplygin gas case $\alpha=1$). The point in the figure in the left indicates
the maximum in the two-dimensional distribution.
}}
\label{2pmChap}
\end{figure}
\end{center}

Case (ii) is the original Chaplygin gas with $\alpha = 1$ and we have two free parameters as well, $\Omega_{m0}$ and $\bar A$. The results of the statistical analysis are visualized in figure \ref{2pmChap}. In the left panel the two-dimensional PDF for $\Omega_{m0}$ and $\bar A$ and $\alpha$ is seen. The lighter the color the higher the probability. The center panel shows the one-dimensional PDF for $\bar A$. It is maximal near $\bar A = 1$. The right panel depicts the one-dimensional PDF for $\Omega_{m0}$. It is maximal around $\Omega_{m0} \sim 0.95$, a value close to unity, i.e., much larger than the baryon fraction.
According to our criterion, the unified model is clearly disfavored, since the GCG fraction is of the order of $5\%$.

\begin{center}
\begin{figure}[!t]	
\begin{minipage}[t]{0.32\linewidth}
\includegraphics[width=\linewidth]{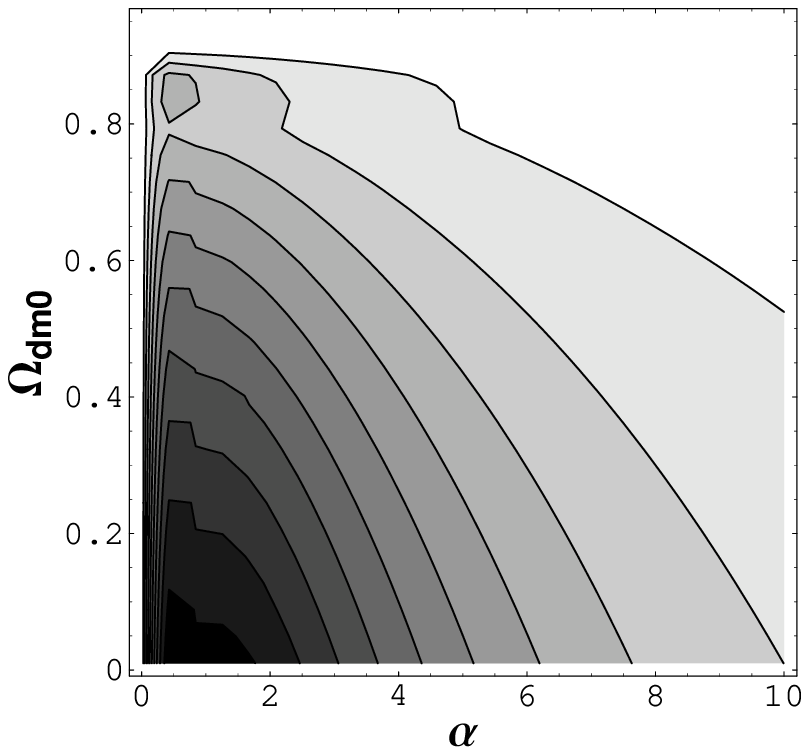}
\end{minipage} \hfill
\begin{minipage}[t]{0.32\linewidth}
\includegraphics[width=\linewidth]{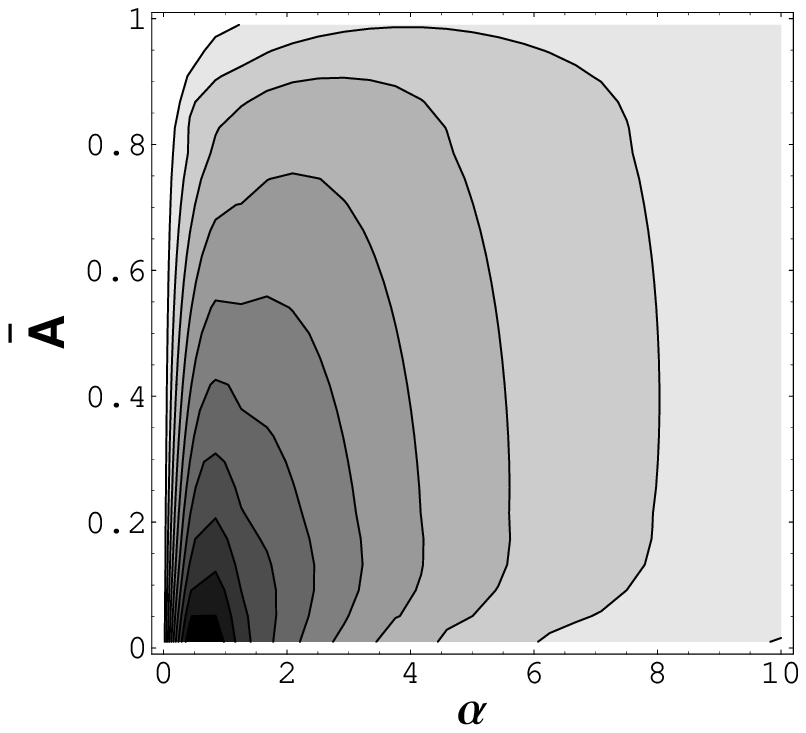}
\end{minipage} \hfill
\begin{minipage}[t]{0.32\linewidth}
\includegraphics[width=\linewidth]{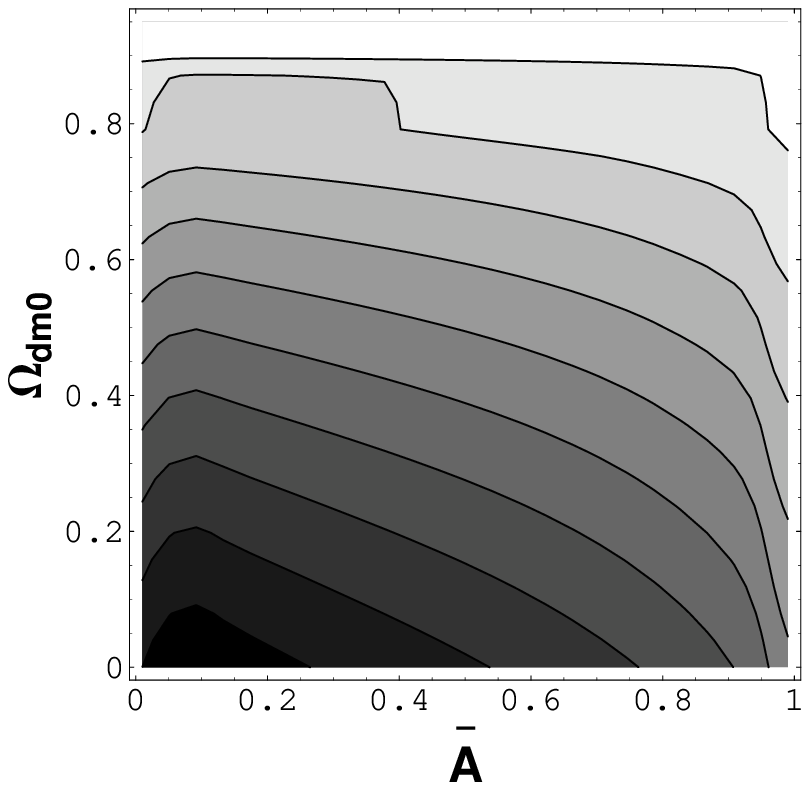}
\end{minipage} \hfill
\caption{{\protect\footnotesize Two-dimensional probability distribution for the different combinations of
the parameters $\alpha$, $\Omega_{dm0}$ and $\bar A$.
}}
\label{PDF2}
\end{figure}
\end{center}

\begin{center}
\begin{figure}[!t]
\begin{minipage}[t]{0.38\linewidth}
\includegraphics[width=\linewidth]{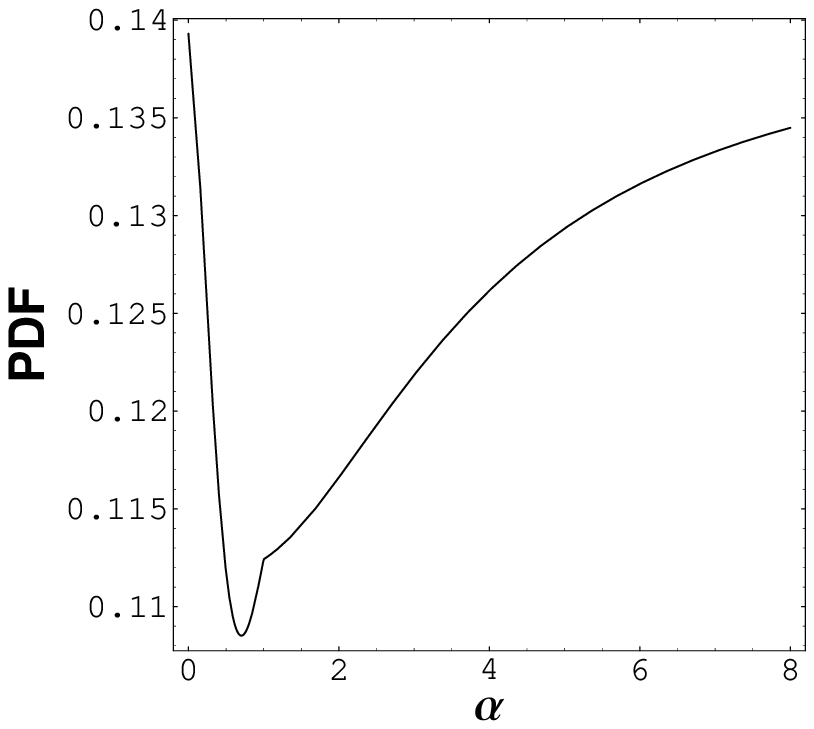}
\end{minipage} \hfill
\begin{minipage}[t]{0.4\linewidth}
\includegraphics[width=\linewidth]{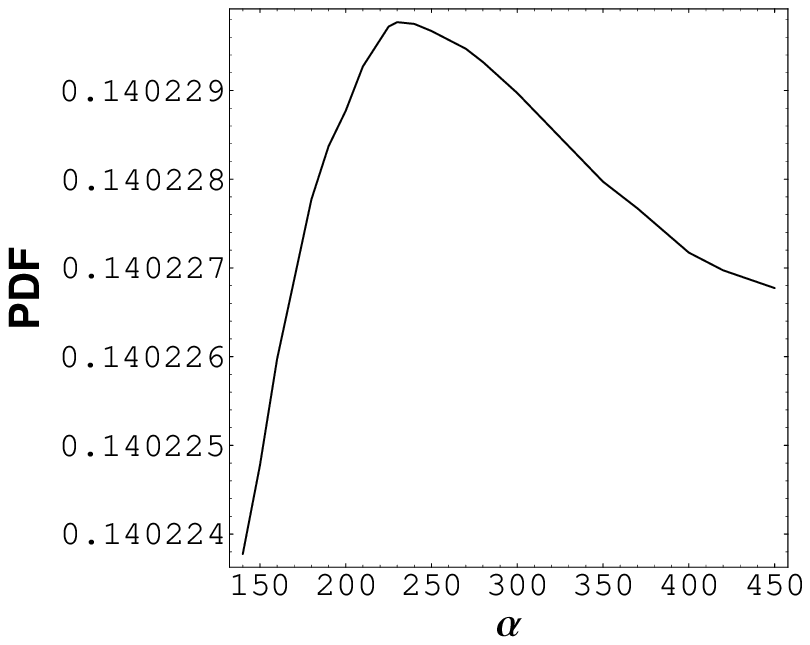}
\end{minipage} \hfill
\begin{minipage}[t]{0.38\linewidth}
\includegraphics[width=\linewidth]{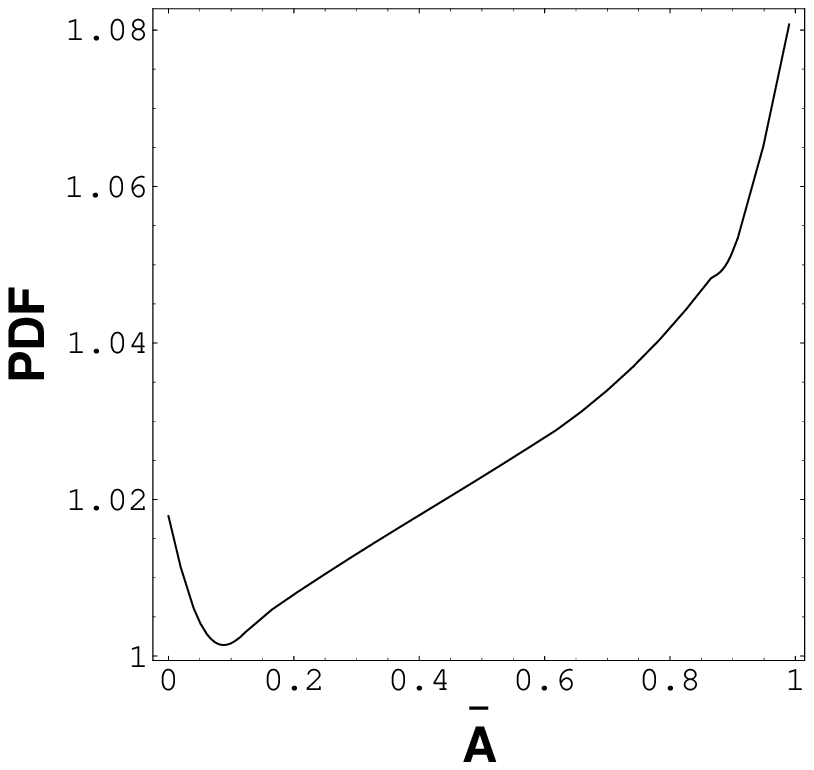}
\end{minipage} \hfill
\begin{minipage}[t]{0.38\linewidth}
\includegraphics[width=\linewidth]{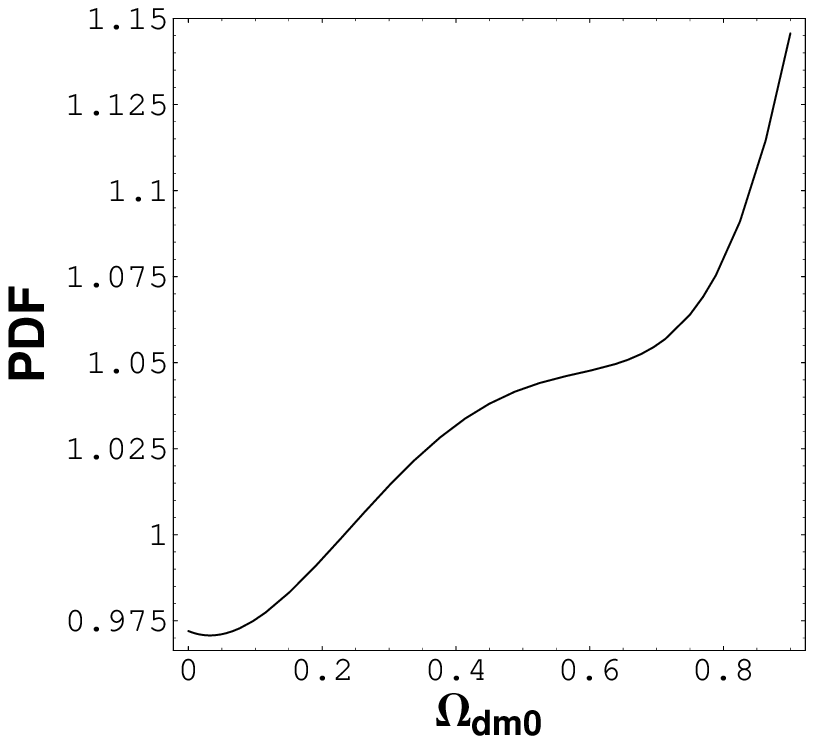}
\end{minipage} \hfill
\caption{{\protect\footnotesize One-dimensional probability distribution for $\alpha$, $\Omega_{dm0}$ and $\bar A$.}}
\label{PDF1}
\end{figure}
\end{center}

In case (iii) all the three parameters $\alpha$, $\Omega_{m0}$ and $\bar A$ are left free. The two-dimensional PDFs for all binary combinations are shown in figure \ref{PDF2}, the one-dimensional PDFs in figure \ref{PDF1}.
Again, the maximum value of $\Omega_{m0}$ is close to unity, confirming the result of case (ii) according to which the contribution of the GCG component is almost negligible.
The behavior of $\alpha$ is similar to that of the unified model of case (i), although this result is less important here since the GCG is subdominant.
\par
We remark that a more detailed, quantitative estimation for the parameters of the model is doubtful due to the existence of different regions
of high probability implying that the one-dimensional PDF is not gaussian.

\section{Conclusions}

Our fully relativistic analysis corroborates the results that were previously obtained within a simplified Newtonian analysis \cite{neo}. As in the Newtonian case, the PDF of $\bar A$ is sensitive to the number of free parameters considered.
For $\alpha$ there are two regions of high probability: one region near zero and a second one for $\alpha \approx 240$. Although the existence of this second region comes as a surprise and represents a new result compared with \cite{neo}, it confirms an entirely independent investigation, using the integrated Sachs-Wolfe effect \cite{piattella}.
The probability distribution for $\Omega_{m0}$ is essentially the same as in the Newtonian case: a universe with negligible
Chaplygin gas component is favored. In fact, our analysis of the large-scale structure data from the 2dFGRS,  taken separately, does not favor an accelerating universe. This is in striking contrast to the results for the homogeneous and isotropic background on the basis of the supernova data \cite{colistete}.
This may imply that the Chaplygin gas model should be discarded. But on the other hand it is well known (see also \cite{neo}), that the matter power spectrum generally does not sufficiently constrain the dark energy component. Further studies, in particular the crossing with other, independent tests are desirable to assess the status of Chaplygin-gas cosmologies.

\par
\noindent
{\bf Acknoweledgments}: We thank CNPq (Brazil) and FAPES (Brazil) for partial financial support. H.E.S.V. thanks for the kind
hospitality of the theoretical physics group at the Universit\"at Bielefeld, Germany, during the elaboration of
this work, and the DAAD (Germany) for financial support.

\end{document}